\documentclass[a4paper,UKenglish]{lipics-v2016}
\usepackage{graphicx,amssymb,amsmath}
\usepackage{microtype}

\title{Approximation Schemes for Covering and Packing in the Streaming Model}

\author[1]{Christopher Liaw}
\author[1]{Paul Liu}
\author[1]{Robert Reiss}
\affil[1]{Department of Computer Science, University of British Columbia, Vancouver, Canada. \texttt{cvliaw@cs.ubc.ca, paul.liu.ubc@gmail.com, rreiss@cs.ubc.ca}}

\Copyright{Christopher Liaw, Paul Liu, and Robert Reiss}


\usepackage{algorithm}
\usepackage[noend]{algpseudocode}
\usepackage{cleveref}
\usepackage{tcolorbox}
\usepackage{tikz}
\usepackage{standalone}
\usepackage{mathtools}
\usepackage[euler]{textgreek}

\usepackage[textsize=scriptsize]{todonotes}

\newcommand{\eps}{\varepsilon}
\newcommand{\E}{\mathbb{E}}
\newcommand{\F}{\mathbb{F}}
\newcommand{\N}{\mathbb{N}}
\newcommand{\R}{\mathbb{R}}
\newcommand{\cA}{\mathcal{A}}

\newcommand{\cH}{\mathcal{H}}
\newcommand{\cO}{\mathcal{O}}
\DeclareMathOperator{\OPT}{OPT}
\DeclareMathOperator{\poly}{poly}

\newcommand{\paren}[1]{\left( #1 \right)}

\begin{document}
\thispagestyle{empty}
\maketitle

\begin{abstract}
 The \emph{shifting strategy}, introduced by Hochbaum and Maass~\cite{Hochbaum1985}, and independently by Baker~\cite{Baker83}, is a unified framework for devising polynomial approximation schemes to NP-Hard problems. This strategy has been used to great success within the computational geometry community in a plethora of different applications; most notably covering, packing, and clustering problems~\cite{Shifting1,Shifting3,Shifting2,Shifting4,Shifting0}. In this paper, we revisit the shifting strategy in the context of the \emph{streaming} model and develop a streaming-friendly shifting strategy. When combined with the \emph{shifting coreset} method introduced by Fonseca et al.~\cite{Fonseca}, we obtain streaming algorithms for various graph properties of unit disc graphs. As a further application, we present novel approximation algorithms and lower bounds for the unit disc cover (UDC) problem in the streaming model, for which currently no algorithms are known.
\end{abstract}

\section{Introduction}
The \emph{shifting strategy} is a unified framework for devising polynomial-time approximation schemes (PTASes) to NP-Hard problems. Originally used by Baker~\cite{Baker83} for maximum independent set in planar graphs, the shifting strategy was modified to solve several geometric covering problems in the widely-cited paper of Hochbaum and Maass~\cite{Hochbaum1985}. Since then, this strategy has found applications in an incredibly diverse set of domains; including facility location, motion planning, image processing, and VLSI design. 

For geometric problems, the shifting strategy is based on partitioning the possible input space into disjoint regions (or windows), solving each disjoint region (either exactly or approximately), and then joining the partial solutions from each window into a candidate global solution. By choosing several partitions, and minimizing over the candidate solutions from each one, a good approximation to the problem is formed. The main observation of the shifting strategy is that the analysis of the approximation factor can be done in two independent portions; the error accumulated from dividing the space into windows, and the error from the within-window algorithm. The within-window algorithm is typically easier to design; in many problems the optimal solution size within a window is bounded by a small constant. Thus by specifying good within-window algorithms, the algorithm designer can get a global solution with only a small overhead in complexity. One of the original problems addressed by Hochbaum and Maass is the \emph{unit disc cover} (UDC) problem: given a point set $P$ in the plane, the problem asks for the size of the smallest set of radius $r$ (or equivalently, unit) discs that cover $P$ completely.\footnote{Actually, Hochbaum and Maass consider the problem of \emph{finding} the smallest of discs that cover a set of points. Our problem is slightly different in that we only care about the \emph{size} of such a cover.} In this case, the partition of the input space is a tiling of the plane by identical $\ell \times \ell$ squares. Within each square, the optimal UDC is found by brute force, as the solution size is at most $\cO(\ell^2)$. By iterating over translates (or shifts) of this tiling, Hochbaum and Maass obtain a $\left(1+\frac{1}{\ell}\right)^{2}$-approximation with running time $n^{\cO\left(\ell^2\right)}$ for UDC in 2D.

Recently, there has been renewed interest in making shifting strategy algorithms practical, as the PTASes obtained by the shifting strategy are too slow to be applied in practice. In recent work by Fonseca et al.~\cite{Fonseca}, the technique of \emph{shifting coresets} is introduced, giving linear time approximations for various problems on unit disc graphs. They observe that within-window algorithms used in the shifting technique often iterate over $m^{\text{poly}(\ell)}$ candidate solutions, where $m$ is the number of points inside the window and $\ell$ is the size of the window. By using coresets to approximate and sparsify point set inside the window, they mitigate the high memory and computational cost of the within-window algorithm. Their algorithms are no longer PTASes, but run in linear time and produce constant factor approximations.

Although the shifting strategy is widely used, scarce attention has been given to it in the \emph{streaming} model. In the streaming model, the complexity of an algorithm is measured mainly by the number of passes it makes over the input data, and the amount of memory used over the duration of the algorithm. In common settings, the requirements are that the algorithm makes only one pass over the input data and uses sublinear (usually polylogarithmic) memory in the size of the input. This is difficult in the context of the shifting strategy as partitioning the input often requires the practitioner to keep a mapping of input points to windows within the partition, necessitating at least linear space. 

In this paper, we revisit the shifting strategy in the context of the streaming model, and develop a streaming-friendly variant. Our streaming shifting strategy only relies on the algorithm designer to design a within-window streaming algorithm $\cA$. Provided that the optimal solution within each window is bounded, the streaming shifting strategy then gives a global algorithm that only introduces a polylogarithmic overhead to the memory use of $\cA$, with the same number of passes over the input data. The analysis is inspired by a recent algorithm of Cabello and P\'{e}rez-Lantero~\cite{CP15}, who presents a $(3/2+\eps)$ approximation for cardinality estimation of maximum independent sets (MIS) of interval graphs in $\cO\left(\text{poly}(1/\eps)\log n\right)$ memory with only one pass over the input data. 

When the memory use of a within-window algorithm for a problem is small (i.e. polylogarithmic), our streaming shifting strategy gives a streaming algorithm for solving the problem globally. Due to this, our results are complementary to those given by Fonseca et al.~\cite{Fonseca}, where $\cO(1)$ memory within-window algorithms are developed for various problems on unit disc graphs. In particular, when their results are combined with our shifting strategy, we obtain streaming algorithms with polylogarithmic memory for independent set, dominating set, and minimum vertex cover on unit disc graphs.

In Section \ref{sec:shifting}, we describe and analyze our streaming shifting strategy. As an application, we present in Section \ref{sec:applications} novel approximation algorithms for the UDC problem in the streaming model. Our UDC algorithms use $\cO\left(\text{poly}(1/\epsilon)\log n\right)$ memory, and operate in only one pass over the input data.  We remark that the results of Cabello and P\'{e}rez-Lantero imply a $(3/2+\epsilon)$ approximation for the 1D UDC problem. This is due to the fact that for unit disc (i.e. unit interval) graphs in 1D, the cardinality of a maximum independent set is equal to the cardinality of a minimum disc cover. However, to the best of our knowledge, UDC has not been considered in the streaming model for 2D and above. In Section \ref{sec:lower-bounds}, we show that any one pass streaming algorithm for 2D UDC in $L_2$ must have approximation factor at least 2.

\section{Preliminaries}
We use the standard notation $[r] = \{1, \ldots, r\}$ where $r \in \N$.
For positive numbers $y, \eps, \delta$, we use the notation $x = y(1\pm \eps) \pm \delta$ to mean $x \in [y(1-\eps) - \delta, y(1+\eps) + \delta]$.
For simplicity, we make the minor assumption that the coordinates of the input points are bounded above by $\poly(n)$ and can be represented using $\cO(\log n)$ bits where $n$ is the number of points.
\subsection{\texorpdfstring{$\eps$}{}-min-wise hashing}
One of the key primitives in our algorithms is the ability to (approximately) sample an element from a set.
To do this, we will use $\eps$-min-wise hash functions which were introduced by Broder~et~al.~\cite{BCFM00}.
We remark that a similar idea was also used in \cite{CP15}.

Let $U = V = \{0, 1, \ldots, k-1\}$ and $\cH$ be a collection of functions $h \colon U \to V$.
We will assume that $k$ is a prime power.
\begin{definition}
	A family of hash functions $\cH$ is said to be $r$-wise independent if for any distinct $x_1, \ldots, x_r \in U$ and any $y_1, \ldots, y_r \in V$ we have
    \[
    	\Pr_{h \in \cH}[h(x_1) = y_1 \wedge \ldots \wedge h(x_{r}) = y_{r}] = \frac{1}{k^r}.
    \]
\end{definition}
Here, we use $\Pr_{h \in \cH}$ to denote the probability measure where each $h$ is drawn uniformly at random from $\cH$.
It is well-known that an $r$-wise independent hash family can be constructed as follows (see \cite{Vad12}).
Let $\F$ be a finite field of size $k$ (such a field exists because $k$ is assumed to be a prime power).
Let $\cH = \{h_{a_0, a_1, \ldots, a_{r-1}} : a_i \in \F\}$ where $h_{a_0, a_1, \ldots, a_{r-1}}(x) = a_{r-1} x^{r-1} + \ldots + a_0$.
Then $\cH$ is an $\ell$-wise independent hash family. Moreover, any element in $\cH$ can be represented using $\cO(r \log k)$ bits.

\begin{definition}
	A family of hash functions $\cH$ is said to be $(\eps, s)$-min-wise independent if for any $X \subseteq [k]$ with $|X| \leq s$ and $x \in X$ we have
    \[
    	\Pr_{h \in \cH}\left[h(x) < \min_{y \in X \setminus \{x\}} h(y)\right] = \frac{1\pm \eps}{|X|}.
    \]
\end{definition}
There is a simple way to obtain $(\eps, s)$-min-wise independent hash functions due to Indyk~\cite{Ind01}.
\begin{theorem}
	\label{thm:min-wise}
	There are fixed constants $c, c' > 1$ such that the following holds.
    Let $\eps > 0$ and $s \leq \eps k / c$.
    Then any $c' \log(1/\eps)$-wise independent hash family $\cH$ is $(s,\eps)$-min-wise independent.
\end{theorem}
For our applications, we will have $s = n$ and $k = \max(n/\eps, \text{poly}(n)^d)$.
In particular, the hash functions can be represented using $\cO(\log^2(1/\eps) + d\log(1/\eps)\log(n))$ bits.
If $\eps^{-1} \leq n$ then this quantity is $\cO(d \log(1/\eps) \log(n))$.

\section{Shifting lemma}
\label{sec:shifting}
We begin by reviewing the shifting strategy of Hochbaum and Maass~\cite{Hochbaum1985} using the UDC problem in $\R^d$ as an example.
For simplicity, we describe the shifting strategy in the planar case $d = 2$.
In the shifting strategy, we partition the plane into windows of size $2 \ell \times 2 \ell$ where $\ell$ is the ``shifting parameter''.\footnote{Hochbaum and Maass~\cite{Hochbaum1985} actually partition the plane into strip of width $\ell$ but small variants, such as replacing strips with windows, also work for identical reasons.}
The windows are closed on the top and left while open on the right and bottom.
We further impose that the coordinates of the top left boundary point are even integers.
Due to these restrictions, there are exactly $\ell^2$ different ways to partition $\R^2$.
Let $S_1, \ldots, S_{\ell^2}$ the be the $\ell^2$ different partitions of the plane.

Suppose that $\cA$ is a within-window algorithm, i.e.~it (approximately) solves the covering problem within a window of size $2 \ell \times 2 \ell$.
Hochbaum and Maass~\cite{Hochbaum1985} proposed the following algorithm to extend $\cA$ to a ``global algorithm'' $\cA_S$.
For each partition $S_i$, we use $\cA$ on each of windows to compute a disc cover.
Then we take the union of the disc cover on each window to produce a global solution $\mathcal{D}_i$.
Having computed $\ell^2$ disc covers, we output the smallest cardinality disc cover of the $\mathcal{D}_i$.
The following lemma states that the approximation ratio of this scheme is not much worse than the approximation ratio of the within-window algorithm.
Hence, to design a global algorithm, one only needs to design a ``local algorithm''.
\begin{lemma}[Shifting lemma~\cite{Hochbaum1985}]
	\label{lem:shift2}
    \[
    	r_{\cA_S} \leq \left( 1 + \frac{1}{\ell} \right)^2 r_{\cA}.
    \]
    where $r_{\cA}, r_{\cA_S}$ are the approximation ratios of $\cA, \cA_S$, respectively.
\end{lemma}
In general, let $\cA$ be an algorithm that approximately solves the disc cover problem in $\R^d$ but restricted to ``windows'' of size $\underbrace{2\ell \times \ldots \times 2\ell}_{\text{$d$ times}}$. Define $\cA_S$ to be the algorithm that partitions $\R^d$ into these windows, uses $\cA$ on each window to find a cover, then takes the smallest cover over all partitions.
Then we have $r_{\cA_S} \leq \left( 1 + \frac{1}{\ell} \right)^d r_{\cA}$.
This is particularly elegant since one can focus on obtaining an approximation algorithm assuming bounded input. Once such an algorithm is developed, it can then be extended to an algorithm on the whole space.

To improve the space complexity of some of our streaming algorithms, we can use the following randomized version of the shifting lemma.
Let $\cA_S$ be the algorithm which randomly picks one of the $\ell^d$ partitions of the $\R^d$ as defined above, say $S_i$, uses $\cA$ to compute a disc cover on each window, then outputs the union as a global disc cover.
\begin{lemma}
	\label{lem:shiftr}
	Suppose $\ell \geq 2d$.
	Then with probability at least $1/2$
    \[
    	r_{\cA_S} \leq \paren{1 + \frac{4d}{\ell}} r_{\cA} \leq \paren{1 + \frac{4}{\ell}}^d r_{\cA}
    \]
    where $r_{\cA}, r_{\cA_S}$ are the approximation ratios of $\cA, \cA_S$, respectively. 
\end{lemma}
\begin{proof}
	Let $S$ be a random partition of $\R^d$ into windows of size $2\ell \times \ldots \times 2\ell$.
    Consider an optimal disc cover and construct a new disc cover as follows.
    If a disc is in $k$ windows then the new disc cover will have $k$ copies of the disc, each associated with one of the windows.
    Note that this gives a disc cover for each of the windows.
    
    Let us number the discs in the optimal cover, $1, \,\ldots,\, \OPT$, and let $X_i$ be the number of windows which contain a portion of disc $i$.
    Since $S$ is a random partition, we have that for each coordinate $j \in [d]$, disc $i$ intersects a closed boundary of a window along coordinate $j$ with probability $1/\ell$.
    If this intersection happens along $k$ coordinates then $X_i \leq 2^k$.
    Hence, $\E X_i \leq \sum_{k \geq 0} \binom{d}{k} \paren{\frac{1}{\ell}}^k \paren{\frac{\ell-1}{\ell}}^{d-k} 2^k = (1 + 1/\ell)^d \leq 1 + 2d/\ell$ where the last inequality is because $\ell \geq 2d$.
    
    Let $Y = \sum_{i=1}^{\OPT} X_i$.
    Then $Y$ is an upper bound on the number of disc covers obtained by solving each window optimally.
    Moreover, $\E[Y - \OPT] \leq \OPT \cdot 2d/\ell$, so by Markov's Inequality, $Y - \OPT \leq  4d /\ell \cdot \OPT$ with probability at least $1/2$.
    The lemma now follows since $\cA$ is an $r_{\cA}$-approximate algorithm for each window.
\end{proof}

\subsection{The shifting lemma in the streaming setting}
In this section, we describe the streaming shifting strategy.
For concreteness, we focus on giving a streaming variant of Lemma~\ref{lem:shiftr}.
Let $\cA$ be a streaming algorithm which approximately solves UDC restricted to a window of size $2\ell \times 2\ell$.

We begin with a high level description of how to use the shifting strategy in the streaming setting.
For now, let us fix a partition of $\R^2$ into windows of size $2\ell \times 2\ell$.
The first issue that arrives is that one is no longer allowed to run $\cA$ on all windows as the space would be prohibitive.
To get around this, we use the following trick from \cite{CP15}.
Set $T = 4\ell^2$.
Let $\gamma_t$ be the number of windows for which $\cA$ outputs a disc cover of size at least $t$.
Since there is a trivial cover of size $T$, we can assume that $\gamma_t = 0$ for $t > T$.
Then the cover obtained by running $\cA$ on all windows is exactly $\sum_{t=1}^{T} \gamma_t$.
The first key observation is that $\gamma_1$ can be interpreted as the number of windows that contain at least one point.
In the language of streaming algorithm, this is exactly the distinct elements problem and can be approximated in very little space.\footnote{Given a stream $a_1, \ldots, a_m \in [n]$, the distinct elements problem is to estimate $|\{a_1, \ldots, a_m\}|$.}
The second key observation is that, if we are able to get a random sample of the windows that contain at least one point then
we can get a very good estimate of the quantity $\eta_t \coloneqq \gamma_t / \gamma_1$.
We can do this approximately using min-wise hashing.

We now commence with a more formal treatment of the above ideas.
Again, let us fix a partitioning of $\R^2$ into windows of size $2\ell \times 2\ell$.
First, we can use an algorithm due to Kane,~Nelson,~and~Woodruff~\cite{KNW10} for distinct elements to obtain the following result.

\begin{lemma}
	\label{lem:gamma1}
	Using $\cO(\eps^{-2} + \log(n))$ bits of space, we can obtain an estimate $\hat{\gamma_1} = (1 \pm \eps) \gamma_1$
    with probability at least $0.99$.
\end{lemma}

Next, we describe how to estimate $\eta_t$ for $2 \leq t \leq T$ using min-wise hashing.
\begin{lemma}
	\label{lem:eta}
	Let $\cA$ be a streaming algorithm for the disc cover problem restricted to a window of size $2\ell \times 2\ell$.
	Suppose that $\cA$ uses $s$ bits of space and let $s_h = \cO(\log(1/\eps)\log(n))$.
    Then using $\cO( \eps^{-2} \ell^4 \log(\ell) (s + s_h))$ bits of space, we can obtain an estimate $\hat{\eta_t} = (1 \pm \eps) \eta_t \pm \eps/T$ for all $t \in \{2, \ldots, T\}$ with probability at least $0.99$.
\end{lemma}
\begin{proof}
	Let $\cH$ be a $\cO(\log(1/\eps))$-wise independent family of hash functions.
    The input to the hash functions is a window (there are $\poly(n)$ possible windows) and the output is a number of $[\poly(n)]$.
    By Theorem~\ref{thm:min-wise}, the family $\cH$ is a $(n, \eps)$-min-wise family of hash functions.

	To estimate $\eta_t$ we do the following.
    Let $r \in \N$ be a parameter to be chosen later and $h_1, \ldots, h_r$ be drawn from $\cH$ uniformly and independently at random.
    For each $j \in [r]$, we maintain a window $W_j$ for $h_j$ and a copy of $\cA$ (denoted $\cA_j$) as follows.
    We initialize $W_j$ to be a dummy window with $h_j(W_j) = \infty$.
    Now suppose we receive a point $p$ in the input and let $W$ be the window that $p$ belongs to.
    If $W = W_j$ then we stream $p$ into $\cA_j$.
    On the other hand, if $W \neq W_j$ then we have two caes.
    If $h_j(W) < h_j(W_j)$ then we replace $W_j$ with the new window $W$, reset $\cA_j$, and stream $p$ into $\cA_j$.
    Otherwise, if $h_j(W) > h_j(W_j)$ then we ignore $p$.

    Fix $t \in \{2, \ldots, T\}$ and $j \in [r]$.
    Let $X_j$ be the random variable which is 1 if $\cA_j$ reports that the window minimizing $h_j$ has a disc cover of size at least $t$.
    Otherwise, $X_j = 0$.
    Since $\cH$ is an $(n, \eps)$-min-wise family, it follows that $\E X_j = (1 \pm 2\eps) \eta_t$.
    Now let $\hat{\eta_t} = \frac{1}{r} \sum_{j=1}^r X_j$.
    By Hoeffding's Inequality, we have
    \[
    	\Pr[|\hat{\eta_t} - \E X_j| \geq \eps / T] \leq 2\exp(-2r\eps^2 / T^2).
    \]
    By choosing $r \geq \cO(T^2 \log(T) / \eps^2)$, the above probability is at most $1/(100T)$.
    Hence, by a union bound, we have $\hat{\eta_t} = (1 \pm 2 \eps) \eta_t \pm \eps/T$ for all $t$ with probability at least $0.99$.

	Finally, it remains to analyze the space requirement of this scheme.
	Storing each hash function requires $s_h$ bits of space.
    Hence, storing all $r$ hash function requires $\cO(\eps^{-2} \ell^4 \log(\ell) s_h)$ bits of space.
    Next, we have a copy of $\cA$ for each of the $r$ windows we maintain.
    So this uses an additional $\cO(\eps^{-2} \ell^4 \log(\ell)s)$ bits of space.
    Hence, the total space usage is $\cO( \eps^{-2} \ell^4 \log(\ell) (s + s_h))$ bits.
\end{proof}
We now prove our main theorem in this section.
\begin{theorem}[Streaming shifting lemma]
	\label{thm:stream_shift}
	Let $\cA$ be a streaming algorithm for the disc cover problem restricted to a window of size $2\ell \times 2\ell$ with approximation ratio $r_{\cA}$.
    Suppose that $\cA$ uses $s$ bits of space and let $s_h = \cO(\log(1/\eps) \log(n))$.
    Then there is a streaming algorithm for the disc cover problem with approximation ratio $(1+\eps)(1 + 4/\ell)^2 r_{\cA}$ that uses $\cO(\eps^{-2} \ell^4 \log(\ell) (s + s_h))$ bits of space and has success probability at least $0.99$.
\end{theorem}
\begin{proof}
	Fix a partition of $\R^2$ into $2\ell \times 2\ell$ windows.
    By Lemma~\ref{lem:gamma1}, with probability at least $0.99$ we obtain an estimate $\hat{\gamma_1} = (1 \pm \eps) \gamma_1$.
    By Lemma~\ref{lem:eta}, with probability at least $0.99$ we obtain an estimate $\hat{\eta_t} = (1 \pm \eps) \eta_t \pm \eps/T$.
    Hence,
    \begin{align*}
		\hat{\gamma_t} & = \hat{\eta_t} \hat{\gamma_1} \\
    	& = \left[ (1 \pm \eps) \eta_t \pm \eps/T \right](1 \pm \eps) \gamma_1 \\
    	& = (1 \pm 3\eps)\gamma_t \pm 2\eps \gamma_1/T.
	\end{align*}
    So
    \[
    	\sum_{t = 1}^{T} \hat{\gamma_t}
        = (1 \pm 3 \eps) \sum_{t = 1}^T \gamma_t \pm 2 \eps \gamma_1 = (1 \pm 5\eps) \sum_{t = 1}^T \gamma_t.
    \]
    If $\eps < 1/10$ then $\sum_{t = 1}^T \gamma_t \leq (1 - 5\eps)^{-1} \sum_{t = 1}^T \hat{\gamma_t} \leq (1 + 20\eps) \sum_{t = 1}^T \gamma_t$.
    Replacing $\eps$ with $\eps/20$, we have a $(1 + \eps)$-approximation to the disc cover computed by running $\cA$ on all windows in the partition.
    
    Finally, by Lemma~\ref{lem:shiftr}, using algorithm $\cA$ on all windows gives a $(1 + 4/\ell)^2r_{\cA}$-approximation algorithm with success probability $0.48$. This can be amplified to $0.99$ by running $\cO(1)$ copies of the algorithm in parallel and taking the median.
    
    The space complexity comes from Lemma~\ref{lem:gamma1} and Lemma~\ref{lem:eta}.
\end{proof}

We remark that our strategy is very general.
In fact, a straightforward extension of our strategy yields the following general theorem for unit disc covers in $\R^d$.
\begin{theorem}
	\label{thm:streaming_d}
	Let $\cA$ be a streaming algorithm for the disc cover problem restricted to a window of size $2\ell \times \ldots \times 2\ell$ with approximation ratio $r_{\cA}$.
    Suppose that $\cA$ uses $s$ bits of space and let $s_h = \cO(d\log(1/\eps) \log(n))$.
    Then there is a streaming algorithm for the disc cover problem with approximation ratio $(1+\eps)(1 + 4/\ell)^d r_{\cA}$ that uses $\cO(\eps^{-2} d^{2d+2} \ell^{2d} \log(\ell d) (s + s_h))$ bits of space and has success probability at least $0.99$.
\end{theorem}

In addition, we do not need to restrict ourselves to single-pass streaming algorithms.
Theorem~\ref{thm:stream_shift} holds whether we consider single-pass streaming algorithms or multi-pass streaming algorithms; one simply needs to use the correct streaming algorithm for $\cA$ restricted to each window.

Using a bit more space will allow us to improve slightly on the approximation ratio in Theorem~\ref{thm:streaming_d}.
This is useful when $\ell$ is a small constant.
\begin{theorem}
	\label{thm:streaming_d2}
	Let $\cA$ be a streaming algorithm for the disc cover problem restricted to a window of size $2\ell \times \ldots \times 2\ell$ with approximation ratio $r_{\cA}$.
    Suppose that $\cA$ uses $s$ bits of space and let $s_h = \cO(d\log(1/\eps) \log(n))$.
    Then there is a streaming algorithm for the disc cover problem with approximation ratio $(1+\eps)(1 + 1/\ell)^d r_{\cA}$ that uses $\cO(\eps^{-2} d^{2d+2} \ell^{3d} \log(\ell d) (s + s_h))$ bits of space and has success probability at least $0.99$.
\end{theorem}
The proof of Theorem~\ref{thm:streaming_d2} is nearly identical to the proof of Theorem~\ref{thm:streaming_d}.
The only difference is that instead of sampling a random partition, we maintain all partitions.
Thus, the space increases by a factor of $\cO(\ell^d)$ but for the approximation ratio, we can apply Lemma~\ref{lem:shift2} instead of Lemma~\ref{lem:shiftr}.

\section{Applications of the streaming shifting lemma}
\label{sec:applications}
In this Section, we present within-window algorithms for unit disc cover and various problems on unit disc graphs. When combined with the streaming shifting strategy, these within-window algorithms give global streaming algorithms.

\subsection{Unit disc cover in 2D with \texorpdfstring{$L_2$}{} balls}
\begin{figure}
\centering
\includegraphics[width=0.5\textwidth]{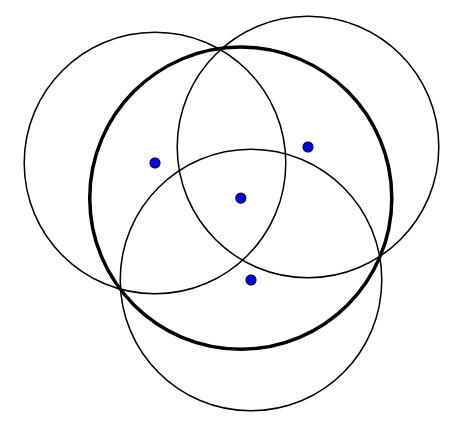}
\caption{A covering of a radius $2/\sqrt{3}$ disc by 3 discs of radius 1. \label{fig:delta-disc}}
\end{figure}

It suffices to give an approximation algorithm for the UDC in 2D restricted to a $2\ell \times 2\ell$ window and then apply Theorem~\ref{thm:stream_shift}.
Let $\delta < \frac{2 / \sqrt{3} -1}{\sqrt 2}$ be a fixed positive constant and partition the window into a uniform grid of side length $\delta \times \delta$.
For each square in the grid, we keep the first point in the stream that lies in the square.
Thus, we only require storing $\cO(\ell^2)$ points and  $\cO(\ell^2 \log(n))$ bits of space for the window.
We then solve the UDC problem optimally given only the points we maintain, giving us a candidate disc cover $\mathcal{C}$.Although $\mathcal{C}$ may not cover all the input points, any uncovered point is at most distance $\delta\sqrt{2}$ from a disc in $\mathcal{C}$. Hence by increasing the radius of each disc in $\mathcal{C}$ by $\delta\sqrt{2}$, we fully cover all the points in the window. By our choice of $\delta$, each disc of radius $1+\delta\sqrt{2}$ can be completely covered by 3 unit discs (see Figure~\ref{fig:delta-disc}), giving a $3$-approximation to the within-window UDC problem.
Choosing $\ell = \cO(1/\eps)$ gives the following theorem.
\begin{theorem}
	There is a streaming algorithm that uses $\cO(\eps^{-8}\log(1/\eps)\log(n))$ bits of space and gives a $(3+\eps)$-approximation to the $L_2$ UDC problem in 2D.
\end{theorem}

We note that the algorithm above can be trivially extended to higher dimensions, though we do not have a good bound on the approximation factor.

\subsection{Unit disc cover in 2D with \texorpdfstring{$L_{\infty}$}{} balls}

Consider as before a $2\ell \times 2\ell$ window.
Recall that an $L_{\infty}$ ball of unit radius corresponds to a $2 \times 2$ square in $\R^2$.
Consider a partition of the window into $\ell$ horizontal strips of unit height. Then this reduces to $\ell$ copies of the standard 1D UDC problem.
We can now use the $(3/2 + \eps)$-approximation for UDC in 1D (due to \cite{CP15}), using $\cO(\eps^{-2} \log(1/\eps) \log(n))$ bits of space for each strip.
Noting that any square in the optimal covering of a
$2\ell \times 2\ell$ window touches at most 2 strips, this gives a $(3 + \eps)$-approximation to the UDC. 
Choosing $\ell = \cO(1/\eps)$ gives a space complexity of $\cO(\eps^{-3} \log(1/\eps) \log(n))$ bits as we require $\ell$ runs of the 1D UDC approximation.
Applying Theorem~\ref{thm:stream_shift} gives the following theorem.

\begin{theorem}
	There is a streaming algorithm that uses $\cO(\eps^{-9}\log^2(1/\eps)\log(n))$ bits of space and gives a $(3+\eps)$-approximation to the $L_{\infty}$ UDC problem in 2D.
\end{theorem}

\subsection{Streaming algorithms for unit disc graphs}
Using the shifting coresets developed in Fonseca et al.~\cite{Fonseca}, we obtain several streaming algorithms for unit disc graphs. In their work, they develop various $\cO(1)$ memory within-window algorithms by computing a coreset for each window. Their coresets are similar to our within-window algorithm for UDC, in that they partition the window into squares of size $\delta \times \delta$ where $\delta$ is a fixed constant. A constant number of points is then stored in each square, and the problem is solved on the stored points. In the offline model, this gives rise to constant factor approximations for maximum weight independent set, dominating set, and minimum vertex cover on unit disc graphs.

Using the streaming shifting lemma, we obtain streaming algorithms for dominating set, minimum vertex cover, and \emph{unweighted} maximum independent set. This is simply from using their within-window algorithms as a black box. The restriction to unweighted problems is due to our technique of subsampling windows, as subsampling may miss a small number of windows that contain large weights of the optimal solution. 

\section{Lower bounds}
\label{sec:lower-bounds}
In this section, we prove lower bounds on the UDC problem.
As is common in streaming lower bounds, our reduction will be via communication complexity.
In particular, we will reduce from the problem \textsc{Index} which is defined as follows.
Let $n \in \N$.
Alice has a vector $x \in \{0,1\}^n$ and Bob has an index $i \in [n]$.
In the one-way communication model, Alice is allowed to send a single message to Bob and Bob must then compute the answer.
Note that there is a trivial protocol that communicates $n$ bits; Alice could send the whole vector $x$ to Bob.
The following theorem asserts that, up to constant factors, there is no better protocol even if it is randomized.
\begin{theorem}[\cite{KNR99}]
	\label{thm:index}
	Any one-way randomized communication protocol which solves \textsc{Index} with probability at least $0.51$ requires $\Omega(n)$ bits of communication.
\end{theorem}

\subsection{A \texorpdfstring{$(1.5-\eps)$}{} lower bound for \texorpdfstring{$L_p$}{} UDC}
In \cite{CP15}, they show that any streaming algorithm that computes a $(1.5-\eps)$-approximation
to the maximum independent intervals problem in one dimension requires $\Omega(n)$ space.
This essentially implies the same lower bound for UDC in any dimension.
\begin{theorem}[\cite{CP15}]
	Fix $\eps \in (0, 0.5)$.
    In all dimensions and for any $L_p$ norm, if a streaming algorithm computes a $(1.5 - \eps)$-approximation to UDC with success probability at least $0.51$ then it uses $\Omega(n)$ space.
\end{theorem}

\subsection{A \texorpdfstring{$(2-\eps)$}{} lower bound for \texorpdfstring{$L_2$}{} UDC in 2D}
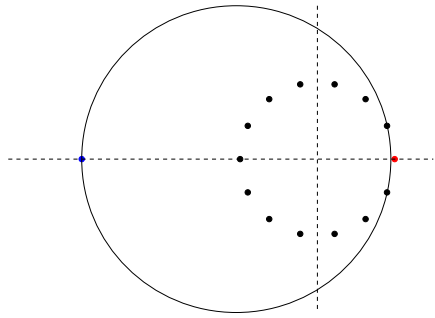
\begin{figure}
  \begin{center}
  \begin{tikzpicture}
    \pgfmathsetmacro{\n}{14}
    \pgfmathsetmacro{\nn}{{\n-1}}
    \foreach \x in {1,...,\nn} {
        \filldraw[black] ({2*cos(\x*360/\n)}, {2*sin(\x*360/\n)}) circle(2pt);
    }
    \filldraw[red] (2,0) circle(2pt);
    \filldraw[blue] ({2*((1+cos(360/\n))/2 - 4)},0) circle(2pt);
    \draw ({2*((1+cos(360/\n))/2 - 2)},0) circle (4);
    
    \draw [dashed] (-8,0) -- (3,0);
    \draw [dashed] (0,4) -- (0,-4);
\end{tikzpicture}
  \end{center}
  \caption{The lower bound construction for UDC in 2D. Alice streams in the points on the unit circle on the right. Bob streams in the point on the left to determine whether or not the rightmost point is present.}
  \label{fig:lb-udc}
\end{figure}

\begin{theorem}
	\label{thm:2d-bound}
	Fix $\eps \in (0,1)$.
    In dimensions two and higher, if a streaming algorithm computes a $(2-\eps)$-approximation to UDC using $L_2$ balls with success probability at least $0.51$ then it uses $\Omega(n)$ space.
\end{theorem}
\begin{proof}
	We will reduce from \textsc{Index}.
    Let $\cA$ be a streaming algorithm, using $S$ bits, which computes a $(2 - \eps)$-approximation to UDC in 2D with $L_2$ balls of radius 2.
    Let $z \in \{0,1\}^n$ be Alice's input and $i \in [n]$ be Bob's input.
    For simplicity, we assume that Bob's input is $i = n$; it will be apparent how to generalize to any $i$.
    If $z_j = 1$ then Alice streams the point $(\cos(2j\pi/n), \sin(2j\pi/n))$ into $\cA$.
    When she is done she sends the memory contents of $\cA$ to Bob.
    Bob now streams the point $\left( \frac{1 + \cos(2\pi/n)}{2} - 4, 0 \right)$ and queries $\cA$. (See also Figure~\ref{fig:lb-udc}.)
    
    Suppose first that $z_i = 0$.
    Then we claim that placing a radius $2$ ball with center at $\left( \frac{1 + \cos(2\pi/n)}{2} - 2, 0 \right)$ covers all the points.
    Indeed, it clearly covers Bob's point.
    To show that the ball covers all of Alice's points, it suffices to show that the radius 2 ball intersects the unit ball for some coordinate in $\left( \cos(2\pi/n), \frac{1 + \cos(2\pi/n)}{2} \right)$.
    Indeed, at $x = \cos(2\pi/n)$, the $y$-coordinates of the radius 2 ball is at $\pm \sqrt{4 - \left( \frac{3 - \cos(2\pi/n)}{2} \right)^2}$.
    It can be verified that the absolute value of this quantity is at least $\sin(2\pi/n)$.
    Indeed, for any $\theta \in \R$
    \begin{align*}
    	& 4 - \left( \frac{3 - \cos(\theta)}{2} \right)^2 - \sin^2(\theta) \\
        & = \frac{3}{4}\cos^2(x) - \frac{3}{2} \cos(\theta) + 3/4 \\
        & = 3 \left(\frac{\cos(\theta) - 1}{2}\right)^2 \\
        & = 3 \sin^4(\theta/2) > 0,
    \end{align*}
    where in the last equality we used the identity $\sin^2(\theta/2) = (1-\cos(\theta)) / 2$.
    Hence, the radius 2 ball covers all of Alice's points so $\cA$ will report a quantity $\leq 2-\eps$.
    
    On the other hand, if $z_i = 1$ then at least two points are required just to cover $(1,0)$ and
    $\left( \frac{1 + \cos(2\pi/n)}{2} - 4, 0 \right)$ so $\cA$ will report $\geq 2$.
\end{proof}

\section{Practical algorithms for UDC}
Although the algorithms of the previous section have low approximation ratios, they involve high constant factors in their running times or memory that may make them unsuitable for practical use. In this section, we develop several streaming algorithms for unit disc cover that we believe are suitable in practice. To achieve good performance in practice, we either relax the approximation factor, or use multiple passes. 

Our first algorithm for UDC is also the simplest. We cover $\mathbb{R}^d$ with an appropriate lattice of unit balls, and then apply the distinct elements algorithm of Kane, Woodruff, and Nelson~\cite{KNW10} to count the number of balls of the lattice containing at least one input point. In the case of $L_\infty$ in 2D, this lattice is simply a uniform grid where each square has width 2. In the case of $L_2$ in 2D, the lattice takes the uniform grid of $L_\infty$ and places a unit circle on each grid point, as well as a unit circle in the center of each grid square. When a point is streamed, we compute the unit ball it belongs to and add that ball to the distinct elements data structure. If the point belongs to multiple balls (as in the $L_2$ case), choose any of the balls it belongs to and add it to the data structure. By choosing randomly from a family of shifted versions of such lattices, we obtain the results below.
\begin{theorem}
\label{thm:simple-udcl2}
There is a one pass streaming algorithm for $L_2$ UDC in 2D that uses $\cO(\eps^{-2}+\log(n))$ space with approximation factor $2\pi(1 + \eps)$ and succeeds with probability at least 0.99.
\end{theorem}

\begin{proof}
Let $\mathcal{S}_{\OPT}$ be the set of discs in an optimal solution. Let $\Gamma$ be the lattice of unit discs described above, and let $n_\Gamma$ be the maximum number of lattice discs intersecting a disc in $\mathcal{S}_{\OPT}$. In the worst case, the algorithm above counts $n_\Gamma$ discs for each disc of $\mathcal{S}_{\OPT}$. 

To compute the expectation from choosing a random shift of the lattice, we can view each disc of $\mathcal{S}_{\OPT}$ as radius 2 and the discs on the lattice as having radius 0 with lattice points on a uniform grid of side length $\sqrt{2}$. Thus $n_{\Gamma}$ is equivalent to the expected number of lattice points that fall within a randomly placed radius 2 disc on the plane. In expectation, this is equal to the area of the disc scaled by the area of a lattice square. Hence we get that $\E n_{\Gamma} = 2\pi$. For each disc in $\mathcal{S}_{\OPT}$, the number of discs it intersects within the lattice is a probability distribution supported on $\{1,2,\ldots,16\}$. Since the mean of the distribution is $2\pi$, running the algorithm with a randomly shifted lattice will produce at most $2\pi\cdot\OPT$ discs with at least a constant probability. By running multiple copies of the algorithm and taking the minimum, we get the result of Theorem~\ref{thm:simple-udcl2}.
\end{proof}

\begin{theorem}
\label{thm:simple-udcl1linf}
There is a one pass streaming algorithm for $L_\infty$ and $L_1$ UDC in 2D that uses $\cO(\eps^{-2}+\log(n))$ space with approximation factor 4 and succeeds with probability at least 0.99.
\end{theorem}

\begin{proof}
The proof of this is exactly analogous to the proof of Theorem~\ref{thm:simple-udcl2} but in this case it is not necessary to randomly shift the lattice. In $L_\infty$ and $L_1$, we use squares instead of $L_2$ discs.
\end{proof}

\subsection{Using multiple passes}
By using multiple passes over the input data, we can give alternate algorithms that both improve the approximation factor and the memory of Theorems \ref{thm:simple-udcl2} and \ref{thm:simple-udcl1linf}.

The following algorithm produces better approximations for UDC in $L_1$ and $L_\infty$ with two passes over the input data. Set the shifting parameter $\ell$ of Theorem~\ref{thm:streaming_d2} to be $2$. For analysis, fix a window and consider the points that fall within the window.
In the first pass, the algorithm goes through the input points and maintains the smallest bounding rectangle that covers all the points.
Observe that we can cover the points with 0 unit squares if and only if the input is empty and we can cover the points with 1 unit square if and only if the bounding rectangle fits inside a unit square.
In either of these two cases, the second pass is not necessary.
If the input points can be covered by 2 unit squares then this can be done by choosing 2 of the 4 corners of the bounding rectangle
and choosing the unit squares to lie in the rectangle while covering the 2 corners.
There are 6 possible ways to do this so in the second pass, we check if one these choices cover all the points.
If not then the point set requires at least 3 squares to cover so we estimate it as 4.
Hence, this gives a $4/3$-approximation for each window.
Combining this with the $9/4$-approximation from using Theorem~\ref{thm:streaming_d2} with $\ell = 2$ gives the following theorem.
\begin{theorem}
There is a two pass streaming algorithm for $L_\infty$ and $L_1$ UDC in 2D that uses $\cO(\eps^{-2}\log n)$ space with approximation factor 3 and succeeds with probability at least 0.99.
\end{theorem}

Finally, we give one additional algorithm for $L_1$ and $L_{\infty}$ UDC in $\R^2$.
Observe that for the 1-dimensional UDC problem, if we allow $\ell$ passes through the data then $O(\eps^{-1}\log n)$ memory suffices to solve the problem with approximation factor $1+\eps$. Within each 1D window, we simply cover the leftmost uncovered point with an interval that begins at that point. By the end of a pass over the input data, we should be able to determine another leftmost uncovered point in the window or if we have covered all of the points. Since all the intervals used are disjoint, we use at most $\ell$ passes for a window size of $\ell$. This effectively simulates the greedy offline interval covering algorithm using multiple passes. Combining this with our streaming strategy gives the following result for $L_{\infty}$ UDC.
\begin{theorem}
There is a $1/\eps$ pass streaming algorithm for $L_\infty$ and $L_1$ UDC in 2D that uses $\cO(\eps^{-7}\log(1/\eps)\log(n))$ space with approximation factor $2+\epsilon$.
\end{theorem}

\begin{proof}
We simply divide each $2\ell \times 2\ell$ window into $\ell$ horizontal strips, and use the 1D UDC algorithm with approximation factor $1+1/\ell$ on each strip for the within-window algorithm. Since each disc of the optimal solution can touch at most two strips, we get approximation factor $2+\epsilon$ by choosing $\ell = \cO(1/\eps)$.
\end{proof}





\small
\bibliographystyle{plainurl}
\bibliography{refs}


\end{document}